\documentclass[aps,pra,twocolumn,floatfix]{revtex4}
\usepackage{graphicx}
\usepackage{bm}
\usepackage{lscape,graphicx}
\usepackage[dvips]{epsfig}
\usepackage{amsmath}

\begin{document}
\title{Local phase space control and
interplay of classical and quantum effects in 
dissociation of a driven Morse oscillator}
\author{Astha Sethi and Srihari Keshavamurthy}
\affiliation{Department of Chemistry, Indian Institute of Technology,
Kanpur, India 208 016}

\begin{abstract}
This work explores the possibility of 
controlling the dissociation of a monochromatically 
driven one-dimensional Morse oscillator by recreating 
barriers, in the form of invariant tori with irrational
winding ratios, at specific locations in the phase space.
The control algorithm proposed by Huang {\it et al.}
(Phys. Rev. A {\bf 74}, 053408 (2006)) is
used to obtain an analytic expression for the control field.
We show that the control term, approximated as an additional
weaker field, is efficient in recreating the
desired tori and suppresses the classical
as well as the quantum dissociation.
However, in the case when the field frequency is tuned close
to a two-photon resonance the local barriers are not
effective in suppressing the dissociation.
We establish that in the on-resonant case quantum
dissociation primarily occurs via resonance-assisted tunneling
and controlling the quantum dynamics requires a local
perturbation of the specific nonlinear resonance in the
underlying phase space.
\end{abstract}

\maketitle

\section{Introduction}
\label{intro}

For over three decades the one-dimensional driven Morse oscillator\cite{morse29} 
has served as a fundamental model to understand and elucidate the 
dissociation mechanism of diatomic molecules.
The continued interest in this, seemingly simple, system is due to two main reasons.
First, the hope is that insights into the mechanism
can be utilized to understand infrared multiphoton
dissociation of polyatomic molecules\cite{lq87,mq89,manz93} and related phenomena including 
vibrational predissociation\cite{bucha93} and mode-specific dynamics\cite{hmm89,crim96}.
Second, at present the focus of researchers is increasingly shifting 
from gaining mechanistic insights to controlling\cite{ts98,bsbook,rice,oskf01} 
the various processes and in this regard a firm understanding of the underlying
mechanisms is essential. 
Therefore, it is not entirely surprising that the driven Morse system has
been studied in great detail from the quantum, classical, and
semiclassical\cite{voth85} perspectives
and with an equally diverse choice for the field - 
monochromatic\cite{w77,d82,wprl86,gm88,gray82,gray84,billi,tann89,breu93,thach95},
bichromatic (with relative phase)\cite{gm882,ns79,bw92,cb95,wu98,nico05,lima08,hcu08}, 
chirped\cite{cbc90,lwy95,kim,feng07}, 
shaped pulses\cite{breu93,cb95,cm93}, and stochastic noise\cite{ken05}.
More recently, the dynamics of a Morse oscillator under the influence of
external fields has become relevant in the context of models for quantum
computing based on molecular vibrations\cite{qvcomp}.

A majority of the studies have addressed the problem from
a classical-quantum correspondence viewpoint; a knowledge of
the regimes where classical or quantum mechanisms are appropriate
and regimes where they coexist and compete is crucial for
control\cite{baybook}. Several important insights have originated from such efforts
which have established that molecular dissociation, in analogy to multiphoton
ionization of atoms\cite{koch85,leo85,Blumel87}, 
occurs due to the system gaining energy by diffusing through the
chaotic regions of the phase space. For example, an important experimental study by 
Dietrich and Corkum has shown\cite{diet92}, amongst other things, the validity of the chaotic
dissociation mechanism.
Thus, the formation of the chaotic regions due to
the overlap\cite{ch79} of nonlinear resonances (field-matter), 
hierarchical structures\cite{percival84,kada} near the 
regular-chaotic borders acting as partial barriers,
and their effects on quantum transport\cite{rad88} 
have been studied in a series of elegant papers\cite{wprl86,y87,grah91}. 
A general consensus, atleast for the one-dimensional driven Morse system,
is that classical-quantum correspondence holds up rather well
except in the regimes of quantum multiphoton
resonances\cite{w77,gm88,gray82,gray84,kim}. 

A recurring theme in many of the works on driven Morse system has to do with enhancing
the dissociation. The search for ways to efficiently dissociate the molecule
has led to a variety of suggestions like bichromatic fields with the relative phase
as a control knob\cite{gm882,ns79,nico05,lima08,hcu08}, frequency-chirped fields\cite{cbc90}, 
and resonant stimulation\cite{kehmm92}. 
However, there are instances wherein one
is interested in suppressing the dissociation rather than enhancing it. This is important,
for example, in the context of vibrational quantum computing\cite{qvcomp} where loss of population
into states other than the states of interest compromises the efficiency of the
quantum gates. Another example comes from coupled Morse oscillator systems where it 
might be neccessary to keep one of the modes
`quiet' in order to carry out mode-specific dynamics\cite{manz93}. A powerful approach to 
implement such constraints on the system comes from 
optimal control theory\cite{rz00} (OCT) and indeed driven Morse
oscillator systems provide an ideal testbed for OCT-based schemes\cite{sr91}.
Yet, in our opinion, it is worthwhile
addressing the issue from a classical-quantum correspondence perspective as well. Not only
is it natural, given the extensive insights that classical mechanics can provide, but
it might also provide a useful way of decoding information buried in an otherwise
complicated optimal field coming out of an OCT calculation. Similar considerations
are at the heart of several works\cite{dynoct} 
aimed at understanding the dynamical origins of the control fields.

Since a detailed understanding of the role of various phase space structures
in the driven Morse system already exists, is it possible to use the phase
space information to control the dissociation using additional, hopefully
simple, fields?
Recently\cite{chandre}, a similar question was addressed by Huang {\it et al.} in the context
of suppressing the multiphoton ionization of atomic systems. 
Using methods\cite{chandre04} developed in a different context,
it was found that the ionization process could be suppressed by rebuilding some
of the broken invariant tori at carefully chosen locations in the phase space.
Inspired by their approach, and noting the mechanistic similarities between
molecular dissociation and atomic ionization\cite{hh92}, 
in this work we attempt to control the 
dissociation of a monochromatically driven Morse oscillator using the local control algorithm.
In their study, Huang {\it et al.} focused\cite{chandre} only on the classical aspects
of suppressing the ionization. It is, however, important to ask if the classical barriers
are effective quantum mechanically as well since
it is not immediately clear that local barriers in the phase space
translate to local suppression of quantum dynamics. We address this issue using the
driven Morse system and show that phase space barriers, especially cantori,
do inhibit both classical and quantum dissociation. As one would expect, such
good classical-quantum correspondence fails in the case of two-photon resonance.
However, we show that the complication comes from a subtle interplay between 
classical and quantum mechanisms with resonance-assisted tunneling\cite{rat0,rat1,rat2} playing
a key role.

We begin by describing some of the salient features of the 
driven Morse oscillator in section~\ref{model}. 
After a brief description of the methodology, section~\ref{dynamics}
contrasts the dissociation dynamics in the off-resonant and on-resonant
situations and a specific initial Morse state is identified to be
subjected to the local control strategy.
In section~\ref{num}, we give a brief summary of the local control method
resulting in an analytic form of the control field. A simplified
control field, appropriate for classical-quantum correspondence studies,
is obtained. 
The efficiency of the simplified control term
in recreating various cantori barriers in phase space and hence controlling the
classical and quantum dissocation dynamics 
is shown and discussed in section~\ref{control}. In the same section we illustrate the
importance of resonance-assisted tunneling in the on-resonance regime.
Finally we conclude in section~\ref{con} with some comments on the method, possible
generalizations, and future applications.  

\section{Model Hamiltonian}
\label{model}

The driven Morse oscillator,
modelling the dissociation
of a diatomic molecule by linearly polarized laser
fields, is described\cite{wprl86} by the Hamiltonian 
\begin{equation}
H(x,p;t)=H_{0}(x,p)-\lambda_1 \mu(x) \cos(\omega_{F} t),
\label{drivmorseham}
\end{equation}	
with the unperturbed Hamiltonian
\begin{equation}
H_{0}(x,p)=\frac{1}{2M}p^{2}+D_{0}{[1-e^{{-\alpha (x-x_e)}}]}^2,
\label{morsepot}
\end{equation}
corresponding to a one-dimensional Morse oscillator.
It is well known that $H_{0}$ provides a good model for
describing the anharmonic vibrations of diatomic molecules with
$D_{0}$, $\alpha$, $x_{e}$, and $M$ being the dissociation energy, 
range of the potential, equilibrium position, and the reduced mass of the
molecule respectively. The bound eigenstates and eigenvalues corresponding to
the Hamiltonian $H_{0}(x,p)$ can
be expressed, with $z \equiv 2ae^{-\alpha(x-x_e)}$, as\cite{mat88}
\begin{subequations}
\begin{eqnarray}
\chi_{\nu}(z)&=&\sqrt{\frac{\alpha (2a-1-2{\nu}){\nu}!}{\Gamma(2a-{\nu})}}\,
   e^{-z/2} z^{b_{\nu}} L_{\nu}^{2b_{\nu}}(z) \label{morseigst}, \\
E_{\nu}&=&
  \frac{2D_{0}}{a}\left({\nu}+\frac{1}{2}\right)-
 \frac{D_{0}}{a^{2}}{\left({\nu}+\frac{1}{2}\right)}^2, 
\label{morseigval}
\end{eqnarray}
\end{subequations}
where $L_{\nu}^{2b_{\nu}}(z)$ is the generalized Laguerre polynomial,
$a=\sqrt{2MD_{0}}/\alpha \hbar$ and,
$b_{\nu}=\sqrt{-2E_{\nu}M}/\alpha \hbar$.

A fit\cite{c81} to the {\it ab-initio} data on HF 
yields the following form for the dipole
\begin{equation}
\mu(x)=A x e^{-\beta {x^4}},
\label{dipo}
\end{equation}
with $A=0.4541$ a.u. and $\beta=0.0064$ a.u.
However, the linear approximation
\begin{eqnarray}
\mu(x) &\approx& \mu(x_{e})+
 \left(\frac{\partial \mu}{\partial x}\right)_{x_{e}}(x-x_{e}) \nonumber \\
&\equiv& \mu(x_{e})+d_{1}(x-x_{e}),
\label{dipo1}
\end{eqnarray}
with $d_{1} \approx 0.33$ a.u. is used in the current work since, 
for the moderate field intensity of interest,
the qualitative nature of classical and quantum dynamics are unaltered as compared to
working with the dipole function in Eq.~\ref{dipo}.
Moreover, as seen later, the linear form allows for a relatively easier implementation of the
local control algorithm~\cite{chandre04} in terms of deriving
analytic expressions for the control field.

Given that this work focuses on suppressing dissociation by creating 
robust Kolmogorov-Arnol'd-Moser (KAM) tori
in the phase space, the action-angle variables $(J,\theta)$ of the unperturbed 
Morse oscillator 
\begin{subequations}
\begin{eqnarray}
J&=&\sqrt{\frac{2MD_0}{\alpha^2}}\left(1-\sqrt{1-E}\right), \\
\theta&=&-sgn(p) \cos^{-1}\left[\frac{1-E}{\sqrt{E}}e^{\alpha(x-x_e)}-
\frac{1}{\sqrt{E}}\right],
\end{eqnarray}
\end{subequations}
are a convenient and natural representation to work with. In terms of $(J,\theta)$
the Hamiltonian in Eq.~\ref{drivmorseham} can be written down as
\begin{equation}
H(J,\theta;t)=H_0(J)-\epsilon v(J,\theta;t),
\label{hamact} 
\end{equation}
with $\epsilon \equiv \lambda_{1} d_{1}$ and
\begin{eqnarray}
H_{0}(J) &=& \omega_0\left(J-\frac{\omega_0}{4D_0} J^2\right), \label{hact0} \\
v(J,\theta;t) &=& 2 \left[V_0(J)
+\sum_{n=1}^{\infty}V_n(J)\cos(n\theta) \right]\cos(\omega_F t), \label{pertact} \nonumber
\end{eqnarray}
being the zeroth-order Hamiltonian and the perturbation respectively.
In the above equations, $E=H_{0}/D_{0}<1$ denotes the dimensionless bound state energy, 
$\omega_0=({2\alpha^2D_0/M})^{1/2}$ is the harmonic frequency, and
$sgn(p)=1$ for $p \geq 0$, $sgn(p)=-1$ for $p < 0$. 
The Fourier coefficients $V_{0}(J)$ and $V_{n}(J)$ 
are known analytically\cite{gm88,grah91} and given by
\begin{subequations}
\begin{eqnarray}
V_{0}(J)&=&\frac{1}{4\alpha}\ln\left[\frac{D_{0}+\sqrt{{D_{0}}^2-D_{0}E(J)}}
{2(D_{0}-E(J))}\right], \\
V_{n}(J)&=&\frac{(-1)^{n+1}}{\alpha n}{\left[\frac{\sqrt{D_{0}E(J)}}{D_{0}+
\sqrt{D_{0}^{2}-D_{0}E(J)}}\right]}^{n}.
\end{eqnarray}
\end{subequations}
Note that the classical nonlinear frequency of the Morse oscillator is given by
\begin{equation}
\Omega_{0}(J)=\frac{\partial H_{0}}{\partial J}=\omega_{0}\left(1-\frac{\omega_{0}}{2D_{0}} J \right).
\label{nonlinfreq}
\end{equation}

Throughout this work, we use atomic units for the
various parameters including the field parameters $(d_{1},\lambda_{1},\omega_{F})$
and measure time in units of the field period $\tau=2\pi/\omega_{F}$. 
In particular, we choose $D_{0}=0.225, \alpha=1.174,
x_{e}=1.7329$ and, $M=1744.59$ corresponding to the Hydrogen Fluoride (HF)
molecule\cite{wprl86}. The potential well supports a total of
$N_{B}=24$ bound states.
The laser field amplitude is fixed at $\lambda_{1}=0.0287$ a.u. ($\sim 30$ TW/cm$^{2}$),
and thus we are studying dissociation under a moderate intensity field. 

\section{Classical and Quantum Dissociation Dynamics}
\label{dynamics}

Although one can choose different classes of initial states for the study, 
in this work the initial
states are chosen to be the zeroth-order Morse eigenstates $\chi_{\nu}$ given
by Eq.~\ref{morseigst}.
The initial states are time evolved on a grid using 
the well established split-operator method\cite{fleck83}
involving the short-time propagator
\begin{equation}
\hat U(\Delta t)=\exp \left (-i {\frac{\Delta t}{2 \hbar}} \hat V \right )
\exp \left (-i {\frac{\Delta t}{\hbar}} \hat T \right )
\exp \left (-i {\frac{\Delta t}{2 \hbar}} \hat V \right ),
\end{equation}
with $T$ and $V$ denoting the kinetic and potential energy operators respectively.
The time-step was set to $\Delta t=5 \times 10^{-3} \tau$   
to ensure convergence of the dissociation probabilities over the timescales of interest
of about $500 \tau$.
As is usual, unphysical reflection at the grid boundaries is avoided by
employing an optical potential\cite{wprl86,lefo83}
\begin{equation}
V_{opt}(x)=-\frac{i V_0}{(1+e^{[-(x-x^{*})/\eta]})},
\end{equation}
with parameters $V_0=0.02$, $\eta=0.35$, and $x^{*}=16.74$.
The introduction of $V_{opt}$ smoothly damps the outgoing wavefunction
and does not modify the time-evolution of the bound states.
The quantum dissociation probability is then calculated as
\begin{equation}
P_{D}^{q}(\tau)=1-\sum_{\kappa=0}^{N_{B}}|\langle \chi_{\kappa} | \chi_{\nu}(\tau) \rangle|^{2},
\end{equation}
where $N_{B}$ is the number of bound states.

In order to compare and contrast the quantum dissociation dynamics with
the classical dissociation dynamics we compute\cite{nico07} the classical dissociation
probabilities $P_{C}(\tau)$ by choosing an
ensemble of initial
trajectories $N_{tot}$ with energy $E_{\nu}$
corresponding to the specific initial Morse state
with the angle uniformly distributed in $[-\pi,\pi]$.
During the time evolution, a trajectory
is considered to be dissociated when the
compensated energy 
\begin{equation}
E_{c} \equiv \frac{1}{2M}{\left [p-\frac{d_{1}\lambda_{1}}{\omega_{F}}
\sin(\omega_{F} \tau)\right ]}^2+D_{0}{[1-e^{-\alpha(x-x_e)}]}^2, 
\label{ecomp}
\end{equation}
exceeds the Morse dissociation energy $D_{0}$.
The number of dissociated trajectories $N_{dis}$ at a given time 
are determined from the above criteria and the resulting classical 
dissociation probability is the fraction
\begin{equation}
P_{D}^{c}(\tau)=\frac{N_{dis}}{N_{tot}}.
\end{equation}

\begin{figure}[t]
\includegraphics[height=80mm,width=80mm]{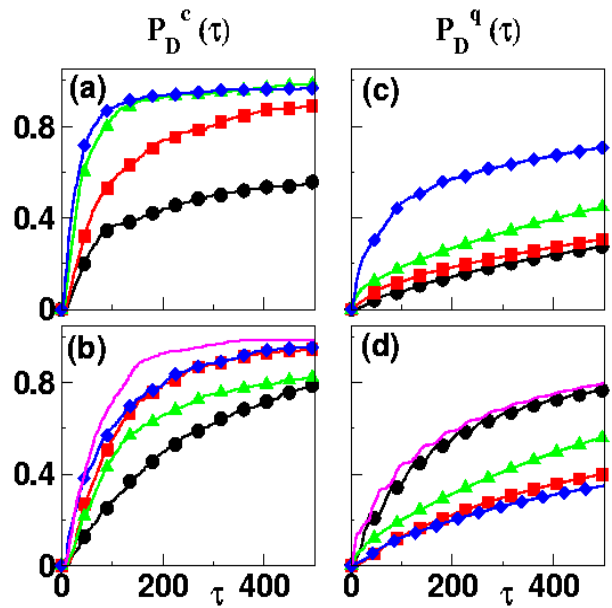}
\caption{(Color online) Classical dissociation probabilities for
the Morse oscillator states $\nu=10$ (circles), $11$ (squares),
$12$ (triangles), and $13$ (diamonds) with driving field frequency
(a) $\omega_{F}=0.0129$ (b) $0.0178$ and fixed field amplitude
$\lambda=0.0287$ ($\sim30$ TW/cm$^{2}$). The corresponding quantum
results are shown in the right column as (c) and (d) respectively. In case of
$\omega_{F}=0.0178$ an additional state $\nu=14$ (line, no symbol) is also shown.
Note that in this figure and all subsequent figures the various parameter values
are in atomic units.}
\label{fig1}
\end{figure}

\begin{figure}[t]
\includegraphics[height=95mm,width=80mm]{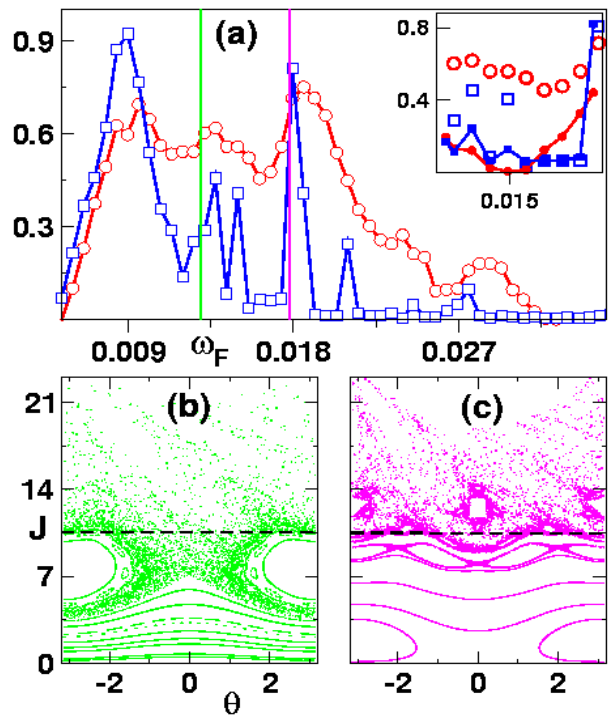}
\caption{(Color online) (a) Quantum (squares) and classical (circles)
dissociation probability for the Morse eigenstate
$\nu=10$ of HF as a function of the field frequency $\omega_{F}$ at the final
time $500 \tau$ with fixed $\lambda_{1}=0.0287$. 
Panels (b) and (c) show the stroboscopic surface of section for two representative cases
considered in this work with $\omega_{F}=0.0129$ and
$0.0178$ respectively. The dashed black 
lines in (b) and (c) are at the classical action value $J=10.5$. 
The inset in (a) shows the effect of building the $1+\gamma^{-1}$ cantorus
barrier on the classical (filled circles) 
and quantum (filled squares) dissociation probabilities.} 
\label{fig2}
\end{figure}

In Fig.~\ref{fig1} we show $P_{D}^{c}$ and $P_{D}^{q}$ as a function of time for
some of the high-lying Morse eigenstates for two specific driving field
frequencies of $\omega_{F}=0.0129$ and $0.0178$. For the specific Morse
parameters and field strength, the lower $\omega_{F}$ represents an off-resonant
situation whereas the higher field frequency corresponds to a two-photon
resonant case. These cases, which will be used to highlight the results, are
selected since they represent two limits in which classical-quantum correspondence
either holds (off-resonant) or does not hold (on-resonant). 
A comparison of the classical (Fig.~\ref{fig1}(a))
and quantum (Fig.~\ref{fig1}(c)) results in the off-resonant case  
reveals that the dissociation probabilities monotonically increase with
increasing vibrational excitation. However, $P_{D}^{q}$ is considerably smaller
as compared to $P_{D}^{c}$. The reasons for this are well known and can be
explained based on the classical phase space shown in Fig.~\ref{fig2}(b). Extensive
classical stickiness\cite{nico07} around the initial action $J=10.5$ (corresponding to
the quantum initial state $\nu=10$) leads to the reduced $P_{D}^{c}$ for this state.
At the same time the density variation in the chaotic regions of the phase space
is symptomatic of the existence of partial barriers - in this case corresponding
to a cantorus with $\omega_{F}/\Omega_{0}(J)=1+\gamma^{-1}$ with  
$\gamma \equiv (\sqrt{5}+1)/2 \approx 1.618$ being the golden mean. 
Based on earlier works\cite{wprl86,grah91},
it is reasonable to surmise that the quantum dissociation is blocked by the cantorus.
On the other hand, results for the on-resonant case shown in Figs.~\ref{fig1}(b) and (d)
indicate a fairly nontrivial behavior. The quantum dissociation probabilities are
non-monotonic with initial states $\nu=10,14$, having nearly identical $P_{D}$, 
dissociating far more than the state $\nu=13$. Quantum mechanically,
resonant two photon transition of state $\nu=10$ to $\nu=14$ leads to
direct coupling with the continuum and hence enhances the dissociation
of state $\nu=10$. The state-to-state
transition probabilities indicate\cite{wprl86}, not shown here, Rabi cycling
between the states $\nu=10,12$, and $14$. Insights into such behavior can also
be gained by studying the classical phase space structures as seen in Fig.~\ref{fig2}(c).
In this on-resonant case $J=10.5$ is essentially
located around the $1+\gamma^{-1}$ cantorus and a prominent $\omega_{F}/\Omega_{0}(J)=2/1$
nonlinear resonance is observed. Clearly, the $2$:$1$ resonance is the classical analog
of the quantum two-photon resonance and must be playing a crucial role in the 
observed dissociation dynamics\cite{gray82}. 
In the subsequent sections we will highlight the classical-quantum correspondence
for both the off-resonant and on-resonant cases.
 
In order to illustrate the key features of this work we focus on the
dissociation dynamics of the Morse state $\nu=10$ for the above mentioned
field frequencies. The analysis, however, can be performed for any initial state
and our specific choice is inspired by the earlier work of Brown and Wyatt\cite{wprl86}.
Moreover, for the HF molecule, under moderate intensity fields, state $\nu=10$ is
a good choice to illustrate the interplay between classical and quantum dissociation mechanisms.
Figure~\ref{fig2}(a) provides the comparison of
quantum and classical dissociation 
probabilities for $\nu=10$ interacting with a field with
fixed intensity and for varying choice of the field frequencies $\omega_F$.
The quantum distribution exhibits peaks at
certain frequencies corresponding to resonant multiphoton transitions.
The classical dissociation profile rises with $\omega_F$, broadens
and dies out smoothly at higher frequencies due
to the transition from trapping of trajectories
in KAM tori at low frequencies to trapping inside the
resonance island regions at higher frequencies.
These observations are rather general and a detailed interpretation
has been given earlier by Nicolaides and coworkers\cite{nico07}.

We now pose several questions in the context of Fig.~\ref{fig2}.
Is it possible to correlate the changes in classical phase space structures 
with varying $\omega_{F}$ with the quantum dissociation probabilities in both
off-resonant and on-resonant cases?
What is the role, if any, of
the classical non-linear resonances in regulating
the decay of quantum states?
Finally, and the main focus of this work, can one control  
the classical and quantum dissociation
dynamics by creating suitable local barriers in the classical phase space?
For the present system the answer is in the affirmative and, as a preview
to the rest of the paper, the inset to Fig.~\ref{fig2}(a) shows
the suppression of classical and quantum dissociation by locally creating
a cantorus with winding number $\omega_{F}/\Omega_{0}(J) = 1+\gamma^{-1}$.
We now turn to the issue of local phase space control of dissociation
which, as seen later, provides answers to the first two questions posed above as well.
 
\section{Control by rebuilding a KAM torus}
\label{num}

The phase spaces shown in Figs.~\ref{fig2}(b) and (c), and the discussion in the
previous section, suggest that if one can rebuild some of the
irrational tori, such as the $1+\gamma^{-1}$ cantorus, locally in the phase
space then it ought to be possible to suppress the dissociation. 
Given the close parallels between the atomic ionization and the system of
interest to us {\it i.e.,} molecular dissociation, 
we employ the same technique\cite{chandre}, 
based on classical perturbation theory, to obtain an analytic expression 
for the control field in case of the driven Morse oscillator.
Since the technique has been described in considerable detail in the 
earlier works\cite{chandre04,chandre}, in what follows we provide
the main results which are of relevance in the present context.
In addition, note that we use the notation of Huang {\it et al.}
for convenience as well as uniformity.

\subsection{Methodology}
To start with, the nonautonomous Hamiltonian is mapped into an autonomous one
by considering $(t ({\rm mod} 2\pi),E)$ as an additional angle-action pair.
Denoting the action and angle variables by ${\bf A} \equiv (J,E)$ and
${\bm \theta} \equiv (\theta,t)$ we can write the original driven system Hamiltonian as
\begin{equation}
H({\bf A},{\bm \theta})=H_{0}({\bf A}) - \epsilon V({\bf A},{\bm \theta}).
\end{equation}
Note that the unperturbed ($\epsilon=0$) invariant tori labeled by the action
${\bf A}$ correspond to the frequency ${\bm \omega} \equiv \partial H_{0}/\partial {\bf A}
=(\Omega,\omega_{F})$. 
The aim is to rebuild a nonresonant torus ${\bf A}_{0}=(J_{0},0)$,
${\bf k} \cdot {\bm \omega} \neq 0$
with integer ${\bf k}$, which has been destroyed due to the interaction with the field.
This can be done by adding a small control term $f({\bm \theta})$ to $H({\bf A},{\bm \theta})$
yielding the control Hamiltonian
\begin{eqnarray}
H_{c}({\bf A},{\bm \theta})&=&H({\bf A},{\bm \theta})+f({\bm \theta}), \nonumber \\
f({\bm \theta})&=& -H({\bf A}_{0}-\partial_{\bm \theta}\Gamma b({\bm \theta}),{\bm \theta}),
\end{eqnarray}
with $b({\bm \theta}) \equiv H({\bf A}_{0},{\bm \theta}) = 
\sum_{{\bf k}} b_{{\bf k}} e^{i {\bf k} \cdot {\bm \theta}}$ 
and $\Gamma$ being a linear operator defined by
\begin{equation}
\Gamma b({\bm \theta}) = \sum_{{\bf k} \cdot {\bm \omega} \neq 0} 
   \frac{b_{\bf k}}{i {\bf k} \cdot {\bm \omega}} e^{i {\bf k} \cdot {\bm \theta}}.
\end{equation}

Referring to the phase spaces shown in Figs.~\ref{fig2}(b) and (c) it is clear that
the classical action corresponding to the quantum initial state $\nu=10$ is located
between the primary resonances $\omega_{F}$:$\Omega_{0}=1$:$1$ and $2$:$1$. Thus,
in our case the aim is to try and rebuild tori with irrational frequency ratios
between the two resonances. In particular, 
the golden mean tori ($1+\gamma^{-n}$, integer $n$) are of specific interest 
in the driven Morse system\cite{wprl86,grah91}.
In the absence of external fields, such an invariant torus with frequency $\Omega_{r}$ is
located at $J_{r}=(\omega_{0}-\Omega_{r})(2D_{0}/{\omega_{0}}^2)$.
We shift the action $\tilde{J}=J-J_{r}$ to focus on the specific region of the 
phase space and expand the autonomous 
Hamiltonian to second order (exact for the Morse
oscillator) in $\tilde{J}$.
Following the methodology outlined above the control term is obtained as 
\begin{equation}
f(\theta,t) 
= \frac{{\omega_{0}}^2}{4D_{0}}{(\partial_{\theta} \Gamma b(\theta))}^2 
+ \epsilon (2 C_{0}+C_{1}), 
\label{control1}
\end{equation}
where we have denoted
\begin{eqnarray}
C_{0} &\equiv& \sum_{k=1}^{\infty} \frac{{(-1)}^k}{k!}V_{0k}(J_{r})
{(\partial_{\theta} \Gamma b)}^k (\cos \omega_F t), \nonumber \\
C_{1} &\equiv& \sum_{n=1}^{\infty}\left \{\sum_{k=1}^{\infty} 
\frac{(-1)^k}{k!} V_{nk}(J_{r}){(\partial_{\theta}
\Gamma b)}^k \right\} \nonumber \\
&\times& \left[\cos (n\theta+{\omega}_F t)+\cos (n\theta-{\omega}_F t)\right],
\end{eqnarray}
with
\begin{eqnarray}
V_{nk}(J_{r}) &\equiv& \left(\frac{d^{k}}{d \tilde{J}^{k}} 
     V_{n}(\tilde{J}+J_{r})\right)_{\tilde{J}=0}, \nonumber \\
\Gamma \partial_{\theta} b(\theta) &=& \partial_{\theta} \Gamma b(\theta) 
= -\epsilon \sum_{n=1}^{\infty}n V_{n}(J_{r}) \\
&{\times}& \left[\frac{\cos(n\theta+{\omega}_{F} t)}{(n\Omega_{r}+\omega_{F})}+
\frac{\cos(n\theta-{\omega}_{F} t)}{(n\Omega_{r}-\omega_{F})} \right]. \nonumber
\end{eqnarray}

For moderate field intensities {\it i.e.,} small $\epsilon$,
one can work with the $O(\epsilon^{2})$ approximation (leading order) to the control term 
in Eq.~\ref{control1} given by
\begin{eqnarray}
f_{a}(\theta,t) &=&
\frac{{\omega_0}^2}{4D_0}{(\partial_{\theta} \Gamma b)}^2 
-2 \epsilon V_{01} (\partial_{\theta} \Gamma b)
\cos(\omega_{F}t) \nonumber \\
&-& \epsilon \zeta(J,\theta;t)\partial_{\theta} \Gamma b(\theta),
\label{control2}
\end{eqnarray}
and it can be shown that
\begin{eqnarray}
V_{01} &=& \frac{{\omega_{0}}^2 }{8\alpha \Omega_{r} D_{0}}
\left(\frac{2\omega_{0}+\Omega_{r}}{\omega_{0}+\Omega_{r}}\right), \nonumber \\
V_{n1} &=& (-1)^{n+1}\left(\frac{{\omega_{0}}^3 }{2\alpha D_0}\right)
\frac{{(\omega_{0}-\Omega_{r})}^{{\frac{n}{2}}-1}}
{{(\omega_{0}+\Omega_{r})}^{\frac{n}{2}+1}}, \\
\zeta(J,\theta;t) &=& \sum_{n=1}^{\infty} V_{n1}(J_{r})
[(\cos (n\theta+{\omega}_{F} t)+\cos (n\theta-{\omega}_{F} t)]. \nonumber 
\end{eqnarray}
It is important to note that 
the perturbative treatment carried out to derive the control
field $f(\theta,t)$ breaks down when $n \Omega_{r} \approx \omega_F$.
Thus, assuming a nonresonant $\Omega_{r}$, the recreated torus 
to $O(\epsilon)$ is located at
\begin{equation}
J(\theta,t)=J_{r}-\partial_{\theta}\Gamma b(\theta).
\label{rectorus}
\end{equation}

\subsection{Simplifying the control term}
\label{simplify}

The control fields $f(\theta,t)$ and $f_{a}(\theta,t)$ obtained above can be 
used for studying the classical dissociation dynamics. However,
a direct use of the control terms in quantum studies is subtle since
the notion of action-angle variables does not exist except in the semiclassical
limit. Thus, in order to implement the classical control terms for studying their
effect on the quantum dissociation dynamics it is necessary to simplify the form
of the control field. Fortunately, Huang {\it et al.} have
already suggested\cite{chandre} such a simplification and we briefly outline their approach.

The control term, being periodic
in $\theta$ and $t$ has a rich Fourier spectrum. However,
only few of the Fourier components are dominant and
the parameter
\begin{equation}
G_{k_1,k_2}\equiv \frac{|F_{k_1,k_2}|}
{|k_{1} \Omega_{r}+k_{2} \omega_{F}|},
\label{pertimp}
\end{equation}
with $F_{k_1,k_2}$ being the coefficients in the double Fourier expansion
of the control term $f$ or $f_{a}$
is used to identify those dominant 
modes. Note that this implies large amplitude $F_{k_1,k_2}$
and $k_{1} \Omega_{r}+k_{2} \omega_{F} \approx 0$ {\it i.e.,}
the corresponding wavevector is close to being in resonance with
the frequency vector ${\bm \omega}$ of the integrable motion.
Once identified, only the dominant Fourier modes are retained in the
control term. Further simplification, required for quantum studies,
is obtained by mapping a typical dominant term as 
\begin{equation}
F_{k_{1},k_{2}} \cos(k_{1} \theta + k_{2}\omega_{F}t) \rightarrow
\lambda_{k_{1},k_{2}} \cos(k_{2}\omega_{F}t).
\end{equation}
The coefficients $\lambda_{k_{1},k_{2}}$ are determined\cite{chandre} by comparing the 
dominant Fourier mode amplitudes in the original control Hamiltonian
with the corresponding amplitudes in the simplified control Hamiltonian
\begin{equation}
H_{c} = H(J,\theta;t) + \mu(x) \lambda_{k_{1},k_{2}} \cos(k_{2} \omega_{F}t).
\label{qcham}
\end{equation}
If more than one dominant Fourier modes are present then 
they will appear as additional terms in Equation~\ref{qcham}.
For all the results presented in the next few sections
we have used the control Hamiltonian of
the form given above.

\section{Influence of the control field on dissociation dynamics}
\label{control}

We now present our results for the effect of local phase space
barriers on the dissociation dynamics of the Morse state $\nu=10$
for the two representative field frequencies and compare to
the uncontrolled results summarized in Fig.~\ref{fig1} and 
Fig.~\ref{fig2}. As we are interested in understanding the effect
of creating cantori barriers on both the classical and quantum
dynamics we also show, following earlier 
studies on quantum transport through cantori\cite{grr86,rad88,maitra00},
the time-averaged probability
\begin{equation}
P_{\nu,\nu'} = \lim_{T \rightarrow \infty} \frac{1}{T}
             \int_{0}^{T} d\tau |\langle \chi_{\nu'}|\chi_{\nu}(\tau) \rangle|^{2},
\label{crossprob}
\end{equation}
of being in a state $\chi_{\nu'}$ having started in the initial state $\chi_{\nu}$.
In this work $T=500 \tau$ is a sufficiently long time for computing
$P_{\nu,\nu'}$. 
The classical analog of Eq.~\ref{crossprob} is constructed by coarse-graining
the actions {\it i.e.,} the trajectory is considered to
be in the action region $J$ if it is located within a bin of width $0.5$ centered
about $J$. Reasonable variations of the bin width lead to qualitatively similar results
and convergence can be easily checked.
Such a coarse-graining procedure is appropriate for studying the
classical-quantum correspondence of $P_{\nu,\nu'}$.

\subsection{Off-resonant laser field}
\label{offrescase}

\begin{figure}[t]
\includegraphics[height=80mm,width=80mm]{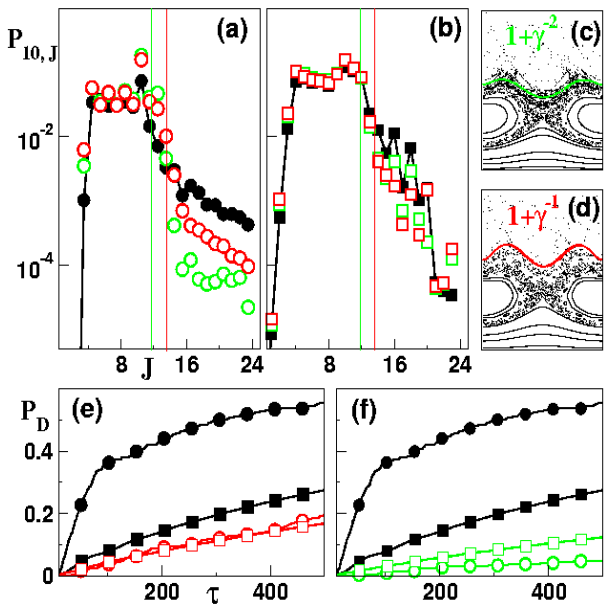}
\caption{(Color online) Effect of creating local barriers
$\omega_{F}/\Omega(J)=1+\gamma^{-1}$ (red, phase space shown in (d))
and $\omega_{F}/\Omega(J)=1+\gamma^{-2}$ (green, phase space shown in (c))
on the classical (circles) and quantum (squares) dissociation dynamics for
$\omega_{F}=0.0129$ (off-resonance). The perturbatively created
tori according to Eq.~\ref{rectorus} are shown in the respective
phase spaces. Panels (a) and (b)
show the classical and quantum time-averaged ($T=500 \tau$) cross probabilities
(cf. Eq.~\ref{crossprob}) respectively with the initial state being the Morse state
$\nu=10$. The thin vertical lines show the expected location of
the KAM barriers.
In (e) and (f) the dissociation probabilities are shown as a function of time
and correspond to (d) and (c) respectively. In the entire figure,
filled and open (red, green) symbols are for the
uncontrolled and controlled cases respectively. See text for discussion.}
\label{fig3}
\end{figure}

Figure~\ref{fig3} summarizes our results for the off-resonant case
with two different cantori barrier being rebuilt in the
phase space. These cantori, corresponding to $\omega_{F}/\Omega_{r}=1+\gamma^{-1}$ (shown
in red), and $1+\gamma^{-2}$ (shown in green), are located at 
actions $J_{r} \approx 13.8$ and
$12.0$ respectively. The modified phase spaces shown in Figs.~\ref{fig3}(c) and (d)
clearly show the reconstruction of the respective barriers as evidenced by the
reduction of stochasticity and increased stickiness around the regular regions. These
phase spaces should be compared to the one shown in Fig.~\ref{fig2}(b) and as
anticipated the control field strength in Eq.~\ref{qcham} $\lambda_{2}$ is indeed
smaller than the driving field strength $\lambda_{1}$. Specifically, 
$\lambda_{2} \approx 0.017$ for the $1+\gamma^{-1}$ barrier and
$\lambda_{2} \approx 0.011$ for the $1+\gamma^{-2}$ barrier with the dominant
Fourier mode being $(k_{1},k_{2})=(3,-2)$ in both cases. In other words the
control field is of the form $\lambda_{2} \cos(2\omega_{F}t)$ and, since
$\lambda_{2} > 0$, comes with a phase difference of $\pi$ relative to the
driving field. Interestingly, Wu {\it et al.} in an earlier work\cite{wu98} have suggested
precisely the same control field characteristics for suppressing chaos
in the driven Morse system. However, they were not clear about the mechanism for
the suppression and this work yields the necessary insight in terms of
the creation of local cantori barriers.

For further insights into the role and efficiency of the cantori barriers
towards controlling the dissociation dynamics, in
Fig.~\ref{fig3}(a) and (b) the classical and quantum time averaged-probabilities
$P_{\nu,\nu'}$ defined in Eq.~\ref{crossprob} are shown. 
Also shown in these figures are the approximate locations of the cantori as thin
vertical lines at the corresponding action values $J=J_{r}$.
The classical $P_{\nu,\nu'}$ in Fig.~\ref{fig3}(a) show that the probabilities fall rapidly
in the vicinity of the rebuilt cantori, especially in case of the
$1+\gamma^{-2}$ cantorus. Consequently, dramatic reduction 
of the classical dissociation probability
in both the cases can be seen (circles) from Figs.~\ref{fig3}(e) and (f).
It is possible to investigate more detailed aspects of the classical 
phase space transport across the
cantori, as done for other systems\cite{grr86,maitra00}, 
but we do not pursue them in this
work. Moreover, it is known that the driven Morse oscillator dynamics near the
separatrix can be analyzed from the perspective of a whisker-map for which Maitra and Heller
have already provided a detailed classical-quantum correspondence of the transport
across cantori\cite{maitra00}.

A key issue that we are interested in this paper is whether the quantum dissociation
is sensitive to the classical phase space barriers being rebuilt. Note that all the
quantum calculations performed herein have $\hbar=1$ and hence we are in the
`quantum regime'. Therefore, {\it a priori} one might anticipate that quantum effects
can override or ignore the changes in the classical phase space.
However, in this off-resonant case, we see from Figs.~\ref{fig3}(e) and (f) that
the quantum results (squares) exhibit clear reduction in the
dissociation probabilities (See inset to Fig.~\ref{fig2}(a) for the entire range
of field frequency). 
Analogous to the classical case, the $1+\gamma^{-2}$ cantorus
is a stronger barrier to dissociation as seen by comparing Fig.~\ref{fig3}(e) with
Fig.~\ref{fig3}(f). The quantum time-averaged probability $P_{\nu,\nu'}$ is shown in
Fig.~\ref{fig3}(b) and exhibits the expected suppression of probabilities for
states lying around and beyond the location of the classical cantori.
Comparing the quantum $P_{\nu,\nu'}$ with the classical results shown in
Fig.~\ref{fig3}(a) we make a few important observations. First, the finite
probabilities for low lying Morse states ($\nu \leq 4$)
seen in the quantum $P_{\nu,\nu'}$ are strictly zero in the classical case.  
This is due to dynamical tunneling through the classical KAM barriers as
proposed nearly two decades ago by Davis and Wyatt\cite{d82}.
Second, the quantum results exhibit oscillations beyond the cantori barrier
in contrast to the smooth classical decay. We suggest that this is a manifestation
of what Maitra and Heller called `retunneling' in their study\cite{maitra00} of the whisker map.
Although the reconstructed cantori are perceived as complete barriers by the quantum
system, some of the quantum states are able to tunnel efficiently across the cantori
since $\hbar$ is large
{\it i.e.,} the quantum mechanism (enhancement due to tunneling) 
dominates the classical (suppression due to cantorus ) mechanism. 
This might explain as to why the suppression of quantum dissociation probability 
due to the $1+\gamma^{-1}$ barrier is not significantly different from that due
to the $1+\gamma^{-2}$ barrier in contrast to the classical results.

Despite the comments made above, it is clear from Fig.~\ref{fig3} that the 
classical-quantum correspondence holds for local phase space control in the
off-resonance case. We now discuss the on-resonant case wherein such a
correspondence is not expected to hold.

\subsection{On-resonance laser field}
\label{onrescase}

As mentioned in the previous section, with the primary driving field frequency
value of $\omega_{F}=0.0178$ the quantum system is in the two-photon
resonant regime involving the Morse states $\nu=10,12$, and $14$. This is
reflected in the quantum dissociation probabilities shown in Fig.~\ref{fig1}(d)
as well as in the classical phase space as a large $2$:$1$ nonlinear
resonance zone (cf. Fig.~\ref{fig2}(b)). Importance of this resonance, embedded in
the chaotic region between the unperturbed $1+\gamma^{-1}$ and $2+\gamma^{-1}$ cantori,
was noted by Brown and Wyatt\cite{wprl86} 
as well as Dardi and Gray\cite{gray82} in an earlier work.
Indeed, our computations (not shown here)
indicate that the Wigner function of state $\nu=12$ is localized on the
resonance with the Wigner functions associated with $\nu=10$ and $14$ straddling
the resonance zone. Such a situation is tailor made for the manifestation of
resonance-assisted tunneling in the system\cite{rat0,rat1,rat2,grah91}.  
Combined with the observation that the initial
state of interest is located right around the $1+\gamma^{-1}$ cantorus (see
Fig.~\ref{fig2}(c)), one expects significant competition 
between the quantum and classical mechanisms for dissociation.
Consequently, the two-photon case provides a difficult challenge for the
local phase space control method.

\begin{figure}[t]
\includegraphics[height=80mm,width=80mm]{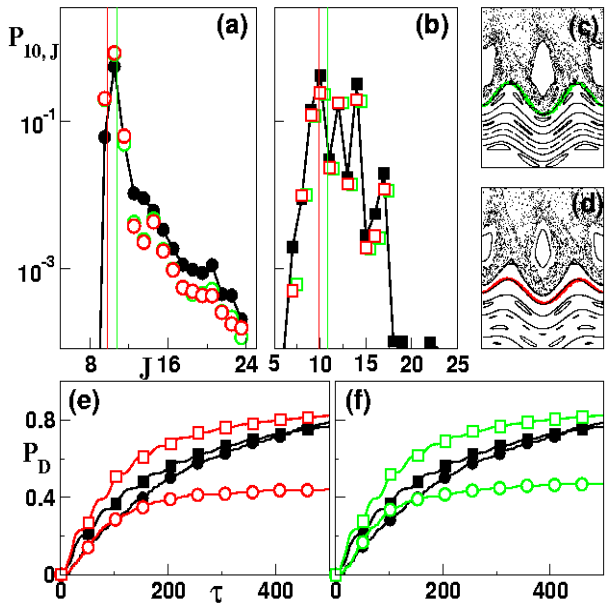}
\caption{(Color online) Results for the on-resonant case $\omega_{F}=0.0178$, 
with notations as in Fig.~\ref{fig3}, 
showing the effect of creating local barriers
$\omega_{F}/\Omega(J) \approx \sqrt{3}$ (red)
and $\omega_{F}/\Omega(J)=1+\gamma^{-1}$ (green).
The respective phase spaces near the $2$:$1$ resonance are shown in (c) and (d).
Note that in this figure the
results corresponding to $\omega_{F}/\Omega(J) \approx \sqrt{3}$
are obtained by retaining the $(3,-2)$ Fourier mode alone. See
the discussion following Eq.~\ref{effl2} and Fig.~\ref{fig5} for
details.}
\label{fig4}
\end{figure}

In Fig.~\ref{fig4} we summarize the results of our attempts to 
control the dissociation by creating the local barriers
with $\omega_{F}/\Omega_{r} \approx \sqrt{3}$ (red) and
$\omega_{F}/\Omega_{r}=1+\gamma^{-1}$ (green). 
We did not attempt to create the 
$1+\gamma^{-2}$ barrier since it would be located much below the
action ($J=10.5$) of the initial state in the classical phase space. 
Results for the two cases will be discussed separately in order to illustrate
the interplay between classical and quantum dissociation mechanisms.
Moreover, in the case of the $\omega_{F}/\Omega_{r} \approx \sqrt{3}$ barrier
complications arise in determining the simplified control Hamiltonian which
requires additional discussion.

Since Fig.~\ref{fig2}(c) shows
that the initial state is located in the region corresponding
to the $1+\gamma^{-1}$ cantorus we attempt to rebuild the cantorus and
Fig.~\ref{fig4}(d) shows that the attempt is successful.
The control field in Eq.~\ref{qcham} is of the form $\lambda_{2} \cos(2\omega_{F}t)$,
corresponding to the dominant Fourier mode $F_{3,-2}$,
with $\lambda_{2} \approx 0.008$ and hence, as in the previous off-resonance
case comes with a relative phase of $\pi$.
Robust barriers have been created in the phase space with a local control field
strength $\lambda_{2}$ which is nearly four times weaker than the driving field strength.
The time-averaged probabilities in Fig.~\ref{fig4}(a) exhibit rapid decay in the
vicinity of the recreated barrier and the classical dissociation probability,
shown in Fig.~\ref{fig4}(d), is reduced by nearly a factor of two.
Surprisingly enough, Fig.~\ref{fig4}(e) shows that the quantum
dissociation is {\em enhanced by a small amount} consistent with the behavior of the
time-averaged probabilities shown in Fig.~\ref{fig4}(b).
The quantum result, in contrast to the off-resonance case, indicates that both
classical and quantum mechanisms are at work in this instance.

\begin{figure}[t]
\includegraphics[height=80mm,width=80mm]{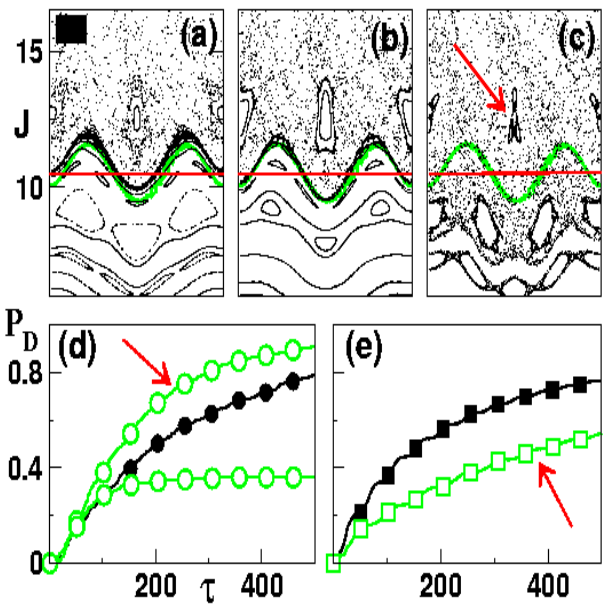}
\caption{(Color online) Phase spaces corresponding to creating
$\omega_{F}/\Omega(J) \approx \sqrt{3}$ barrier 
in the on-resonant case with (a) Full leading
order control term (cf. Eq.~\ref{control2}) (b) Retaining two dominant Fourier modes
(cf. Eq.~\ref{appcontham}) and (c) Simplified control term (cf. Eq.~\ref{qcham})
using both Fourier modes and effective strength $\lambda_{2}$ obtained from
Eq.~\ref{effl2}. In (a) the size of the black square represents $\hbar=1$.
The rebuilt torus can be seen in (a) and (b), highlighted in green, but not
in (c). Note the perturbation of the $2$:$1$ resonance in case (c) indicated
by a red arrow. In (d) and (e) the classical (corresponding to
the figure (b) only) and quantum dissociation probabilities
respectively are shown for the various cases. Curves marked by red arrow correspond
to the phase space (c).}
\label{fig5}
\end{figure}

On the other hand, attempts to create the $\omega_{F}/\Omega_{r} \approx \sqrt{3}$
barrier poses a problem, associated with the simplification of the
control term Eq.~\ref{qcham}, which was not encountered in the previous examples. 
Interestingly, in this case two dominant Fourier modes $F_{3,-2}$ and $F_{4,-2}$ 
are found with the corresponding values $G_{3,-2} \approx 0.029$ and
$G_{4,-2} \approx 0.024$ (cf. Eq.~\ref{pertimp}).
Taking into account only the marginally dominant $(3,-2)$ mode
Fig.~\ref{fig4}(c) shows that the desired local barrier is created.
Consequently, Fig.~\ref{fig4}(f) shows that
the classical dissociation probability is, as in the case
of the $1+\gamma^{-1}$ barrier, reduced by a factor of two.
Again one observes that the quantum counterpart behaves in an opposite manner
{\it i.e.,} the dissociation is slightly enhanced.
However, given that the Fourier mode $(4,-2)$ is nearly as dominant as the
$(3,-2)$ mode, it seems reasonable to use an effective $\lambda_{2}$ in
the simplified control term of Eq.~\ref{qcham} as
\begin{equation}
\lambda_{2} = \frac{F_{3,-2}}{V_{3}(J_{r})} + \frac{F_{4,-2}}{V_{4}(J_{r})}.
\label{effl2}
\end{equation}
Such a procedure yields $\lambda_{2} \approx -0.015$ and thus the control
field, still less intense than the primary field, comes with a relative
phase of zero. Nevertheless, 
such an attempt fails as seen by inspecting 
the phase space shown in Fig.~\ref{fig5}(c) which exhibits
increased stochasticity and, expectedly, leads to enhanced classical dissociation
observed in Fig.~\ref{fig5}(d). Now, however, Fig.~\ref{fig5}(e) shows
that {\em the quantum dissociation
is suppressed appreciably} and hence it would seem as if the quantum dynamics
feels the barrier where there is none! Arguments invoking the large, finite value
of $\hbar$ and dynamical localization can be safely ignored since the
phase space in Fig.~\ref{fig5}(c) does not show any appreciable stickiness near the
apparent barrier. The resolution to such an unexpected result comes from a closer
inspection of the phase space in Fig.~\ref{fig5}(c). One can clearly see that
the attempt to create the local barrier has resulted in a severe perturbation
of the $2$:$1$ nonlinear resonance. As the additional Fourier mode $(4,-2)$
implies $2\Omega_{r} \approx \omega_{F}$, the observed perturbation can
be tracked to the specific mode as long as it comes with an opposite phase.
This will be established in the following subsection (cf. Eq.~\ref{ratdest}).
Since this specific resonance is key
to the two-photon process, it should not be surprising that the quantum dissociation
is suppressed.

It is crucial to note that failure to
create the barrier of interest occurs only when the control term is
mapped to the simplified form as in Eq.~\ref{qcham} using the effective
value for $\lambda_{2}$ shown above. This is confirmed by inspecting the
phase spaces shown in Fig.~\ref{fig5}(a),(b) 
associated with the full $O(\epsilon^{2})$ control Hamiltonian (see Eq.~\ref{control2})
and the approximate Hamiltonian
\begin{eqnarray}
H_{c}(J,\theta,t) &\approx& H(J,\theta,t) \nonumber \\
&+&\sum_{n=3,4}^{} F_{n,-2} \cos(n\theta-2\omega_{F}t),
\label{appcontham}
\end{eqnarray}
obtained by retaining only the dominant Fourier modes in the control term.
The recreated barriers in the phase space can now be clearly seen and
Fig.~\ref{fig5}(d) shows that the classical dissociation computed using
the Hamiltonian in Eq.~\ref{appcontham} is indeed suppressed.
Clearly, the opposing classical and quantum
results in this subsection, with the associated phase space structures, point to
the importance of the quantum dissociation mechanism in the on-resonant case.
In what follows we show that these observations can be rationalized based on the
phenomenon of resonance-assisted tunneling.

\subsubsection{Dissociation via resonance-assisted tunneling}
\label{rat}

In order to confirm the above arguments and to gain a better understanding of
the results shown in Fig.~\ref{fig4} and Fig.~\ref{fig5} for the
dissociation of the state $\nu=10$ we focus on the role of $2$:$1$ resonance
within the paradigm of resonance-assisted tunneling.
Following the earlier works\cite{rat0,rat1,rat2}, the motion in the vicinity of a $r$:$s$ resonance
is analyzed by applying secular perturbation theory  
on the uncontrolled Hamiltonian in Eq.~\ref{hamact} and for details we refer the
reader to the original work\cite{rat0}.
First, a canonical transformation to the appropriate slow angle
$\theta \rightarrow \phi=\theta-\Omega_{r:s}t, 
J \rightarrow J$ is made resulting in the new Hamiltonian
\begin{equation}
\tilde{H}(J,\phi,t) = H_{0}(J) - \Omega_{r:s} J + \tilde{V}(J,\phi,t), 
\label{trancham}
\end{equation}
where $\tilde{V}(J,\phi,t)=V(J,\phi+\Omega_{r:s}t,t)$
and we have denoted $V(J,\theta,t) \equiv -\epsilon v(J,\theta;t)$ (Cf. Eq.~\ref{pertact}). 
We now expand $H_{0}$ in Eq.~\ref{trancham} about the resonant action $J_{r:s}$ 
to obtain the zeroth-order Hamiltonian in the vicinity of $r$:$s$ of the form 
\begin{equation}
\tilde{H}_{0}(J)=\tilde{H}_{0}(J_{r:s})+\frac{1}{2m_{r:s}}{(\Delta J)}^2,
\end{equation}
with $\Delta J \equiv J-J_{r:s}$ and $m_{r:s} \equiv -2D_{0}/\omega_{0}^2$.
Since, $\phi$ varies slowly near $r$:$s$, $\tilde{V}(J,\phi,t)$
is replaced by its average over $r$ field periods 
\begin{eqnarray}
V_{\rm av}(J,\phi) & \equiv & \frac{1}{r\tau} \int_{0}^{r\tau}
\tilde{V}(J,\phi,t)dt \nonumber \\
& \simeq & \lambda_{1} \sum_{n=1}^{\infty} V_{n}(J) \cos(n r\phi).
\end{eqnarray}
Ignoring the higher harmonics in the above expression and neglecting the 
action dependence of the Fourier coefficients $V_{n}$ we 
obtain an effective (integrable) pendulum Hamiltonian  
\begin{equation}
H_{eff}(J,\phi) \simeq \frac{1}{2\tilde{m}_{r:s}}{(\Delta J)}^2 +
2\tilde{V}_{r:s}(J_{r:s}) \cos(r\phi),
\label{pendham}
\end{equation}
to describe the dynamics near the $r$:$s$ resonance with
$\tilde{m}_{r:s}=-m_{r:s}$ and 
$V_{r:s}(J)=\lambda_{1} d_{1} V_{1}(J)/2$.

Specializing the above result to the observed $2$:$1$ resonance,
the resonant action is determined to be $J_{2:1}=12.6$, thus confirming
the participation of the Morse state $\nu=12$. 
Using the zeroth-order Hamiltonian in Eq.~\ref{pendham} one finds,
with $J=10.5$ (quantum state $\nu=10$) and $J'=14.5$ (quantum
state $\nu'=14$), that
\begin{eqnarray}
|E_{J}-E_{J'}|&=&\left|\frac{1}{2\tilde{m}_{2:1}} (J-J')(J+J'-2J_{2:1}) \right| \\
               &\approx& 3.2 \times 10^{-4}, \nonumber 
\end{eqnarray}
{\it i.e.,} the states $\nu=10$ and $\nu=14$ are nearly symmetrical with
respect to the state $\nu=12$ localized on the $2$:$1$ resonance\cite{grah91,rat0}.
Therefore, the nonzero coupling $V_{2:1}$ will efficiently connect the states
$\nu=10$ and $\nu=14$. Moreover, it is possible to estimate the strength of the resonance
for the given parameters as $V_{2:1}(J_{2:1}) \approx 0.01464$.  
It is crucial to note that the the effective control field coupling strength 
from Eq.~\ref{effl2}
in case of $\omega_{F}/\Omega_{r} \approx \sqrt{3}$ {\it i.e.,} relevant to
the phase space shown in Fig.~\ref{fig5}(c) satisfies
\begin{equation}
\lambda_{2} \approx -V_{2:1}(J_{2:1}). 
\label{ratdest}
\end{equation}
Thus the control field with a zero relative phase tends to
negate the effect of the $2$:$1$ resonance generated by the primary
driving field and this can be clearly seen in Fig.~\ref{fig5}(c).
We also remark here that the substantial quantum dissociation probability
seen in Fig.~\ref{fig5}(e), despite the strong perturbation of the
$2$:$1$ resonance, is due to the higher harmonics which have
been neglected in the present analysis.
This confirms our suspicion that the decay of state $\nu=10$ in the two-photon
regime is dominated by resonance-assisted tunneling.

\begin{figure}[t]
\includegraphics[height=80mm,width=80mm]{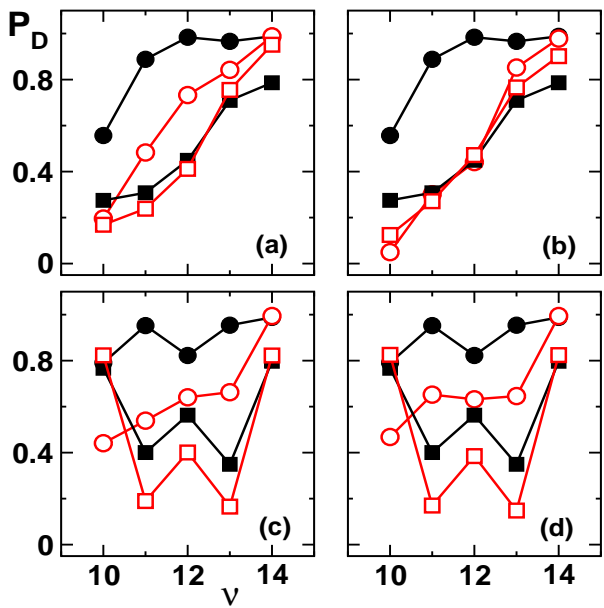}
\caption{(Color online) Effect of the local control term on the dissociation probability
of nearby states, specified by $\nu$, at $t=500 \tau$.
(a) and (b), with off-resonance $\omega_{F}=0.0129$, correspond
to the creation of the $1+\gamma^{-1}$ and $1+\gamma^{-2}$ barriers respectively.
Panels (c) and (d), with on-resonance $\omega_{F}=0.0178$,
correspond to the creation of the $1+\gamma^{-1}$ and $\sqrt{3}$ barriers
respectively.
with fixed field strength as in the previous figures.
The uncontrolled results are shown as filled symbols, classical (circles) and
quantum (squares), while the controlled results are shown as corresponding
open symbols. The lines are drawn as a guide to the eye.}
\label{fig6}
\end{figure}

\subsection{Are the rebuilt barriers local?}
\label{isitlocal}

Up until now most of our results and discussions have focused on a specific
initial Morse state. The barriers were created locally in phase
space to influence the dissociation dynamics of the state $\nu=10$. From
a control point of view, it is of some interest to examine the effect
of such barriers on the dynamics of other states, especially states
that are in the vicinity of state $\nu=10$. In other words, how local
are these barriers? Towards this end,
in Fig.~\ref{fig6} we show the influence of the local barriers
on the dissociation probabilities of other nearby Morse states with $\nu > 10$. 
At this stage it is usefull to recall the results shown in Fig.~\ref{fig1}
with the essential differences between the off-resonant and on-resonant
cases. The effect of the $1+\gamma^{-1}$ and $1+\gamma^{-2}$ barriers
in the off-resonant case, located around $J_{r} \approx 13.8$ and $12.0$,
are shown in Figs.~\ref{fig6}(a) and (b) respectively.
These figures confirm that to a large extent the barriers are indeed local
{\it i.e.,} dissociation is suppressed to varying extent for
initial states lying below the barrier. For states lying above the barrier, 
dissociation is either enhanced (mostly in the quantum case) or slightly
reduced. 

On the other hand the quantum results for the on-resonant case, shown
in Figs.~\ref{fig6}(c) and (d) for the $1+\gamma^{-1}$ ($J_{r} \approx 9.9$)
and $\sqrt{3}$ ($J_{r} \approx 10.9$) barriers (corresponding to
Fig.~\ref{fig4}(c) {\it i.e.,} with only the $(3,-2)$ Fourier mode included)
respectively, are far less straightforward to interpret. This is not 
entirely surprising since, as shown in the last section, resonance-assisted
tunneling plays an important role and overrides the suppression due
to the local barriers. In particular, the dissociation probabilities for
states $\nu=10$ and $\nu=14$ increase very slightly.
However, for the states $\nu=11$ and $\nu=13$, not involved in the resonance-assisted
tunneling process, one observes substantially reduced dissociation despite 
being located far above the local barriers. We suspect that this is due to the
increased stickiness around the $2$:$1$ resonance region observed in the
controlled phase spaces shown in Figs.~\ref{fig4}(c) and (d). Note that this is
corroborated by the observation that the concerned states also exhibit 
reduced classical dissociation as seen in Figs.~\ref{fig6}(c) and (d). 
Further confirmation comes from our calculations which show an opposite
quantum trend to that of Fig.~\ref{fig6}(d) upon inclusion of the
$(4,-2)$ Fourier mode as well resulting in the phase space shown in Fig.~\ref{fig5}(c). 
It is possible to implicate, albeit indirectly, the local barriers with the observed
suppression since creating the barriers leads to a more regular phase space, slightly
increased size of the $2$:$1$ resonance region, and therefore increased stickiness.
Nevertheless, comparing the off-resonant and on-resonant cases shown in Fig.~\ref{fig6},
it is evident that the effects of creating local phase space barriers can be far more
subtle to interpret in the latter case.

\section{Conclusions}
\label{con}

To summarize, this work demonstrates that it is possible to control the
dissociation dynamics of a driven Morse oscillator by creating local
phase space barriers. A clear understanding of the effect of cantori 
on both classical and quantum dissociation dynamics is obtained (see
inset in Fig.~\ref{fig2}(a) for example).
This work also highlights the essential difference between off-resonant
and on-resonant dynamics with resonance-assisted tunneling playing
a prime role in the latter case. Although local phase space barriers
are very efficient in reducing the dissociation in the off-resonant case,
the results in Fig.~\ref{fig4} and Fig.~\ref{fig5} suggest that controlling
the quantum dissociation in the on-resonant regime can be achieved by 
using control fields which selectively perturb the 
appropriate nonlinear field-matter resonance. Similar
observations have been made earlier in a different context\cite{rat2}
and further studies are required to establish a general criteria for controlling
the multiphoton processes from the viewpoint of local modification
of the phase space structures.

Several questions, however, remain to be addressed and we mention a few
important ones. First, there is the issue of the effectiveness of the local barriers in systems
with more than two degrees of freedom since the invariant tori do not
have the correct dimensionality to partition the phase space. However, there
are reasons to hope that even for systems with higher degrees of freedom the
rebuilt tori can act as barriers for short times\cite{remnants}. A careful study of
the classical and quantum transport with and without the control fields
is required in this instance. Apart from polyatomic molecules,
this is also important while considering the
infrared multiphoton dissociation dynamics of a Morse oscillator by
explicitly taking into account the rotations\cite{lmw81,par88}.  
Second, one would like to
extend the approach to systems under the influence of more general time-dependent
fields, for e.g., chirped fields. For slow chirping this should be feasible as
one can then utilize the concept of adiabatic Floquet theory\cite{guerin97}. In case of
arbitrary time-dependent fields the correct approach is not clear at the present moment.
Third, the method used in the present and earlier works is dependent on our ability
to express the Hamiltonian in terms of appropriate action-angle variables. This might
pose some limitations which, as seen in the present work, can be more severe in
terms of implementing the control into the quantum dynamics. Finally, 
for systems with small effective $\hbar$, the resonance-assisted tunneling mechanism
will be replaced by chaos-assisted tunneling\cite{cat} and it would be interesting to
study the influence of the cantori barriers in this context. 
These issues are the focus of
ongoing work in our group.

\section*{Acknowledgments}

It is a pleasure to acknowledge useful discussions
with Prof. Cristel Chandre and Shu Huang 
on various aspects of local control theory. 
Astha Sethi is funded by a Fellowship
from the University Grants Commission, India.

\end{document}